\documentclass[aps,pra,twocolumn,showpacs,amsmath,amssymb]{revtex4-1}

\usepackage{graphicx}
\usepackage{color}
\usepackage{epstopdf}
\usepackage{amsmath}
\allowdisplaybreaks[4]
\usepackage{bm}
\usepackage{hyperref}
\usepackage{cleveref}
\usepackage{tcolorbox}
\usepackage{braket}
\usepackage[export]{adjustbox}
\usepackage[hang]{subfigure}

\begin{document}
	\title{Quantum droplets in dipolar condensate mixtures with arbitrary dipole orientations}
	
	\author{Liang-Jun He}
	\author{Bo Liu}
	\author{Yong-Chang Zhang}
	\email{zhangyc@xjtu.edu.cn}
	\affiliation{MOE Key Laboratory for Nonequilibrium Synthesis and Modulation of Condensed Matter, and Shaanxi Key Laboratory of Quantum Information and Quantum Optoelectronic Devices, School of Physics, Xi’an Jiaotong University, Xi’an 710049, People’s Republic of China }
		
\begin{abstract}
	Through considering a two-component dipolar Bose-Einstein condensate, we investigate the influence of the angle between the polarization orientations of the two species on the ground states, and show that the miscibility between the two components can be adjusted not only by the inter-component contact interaction but also by the polarization angle. Particularly, in the presence of an external confinement, the two species exhibit a preference for splitting into multiple droplets separated from each other in the immiscible regime, featuring a zig-zag like profile. Furthermore, the number of separated droplets depends solely on the polarization angle. This introduces a promising degree of freedom for exploring emergent states of matter in dipolar quantum gases.
	
\end{abstract}

\maketitle

\section{INTRODUCTION}\label{introduction}
In recent years, the dipolar Bose-Einstein condensate (dBEC) has emerged as a promising platform for exploring exotic states of matter led by long-range interactions and quantum fluctuations. Initially, according to mean-field theory, this system was thought to be unstable and prone to collapse due to attractive interactions along the polarization direction~\cite{lahaye2009physics}. However, experimental observations have defied this expectation, revealing that rather than collapsing, dBEC can be stabilized by quantum fluctuations, forming what is known as an emergent self-bound state, namely a quantum droplet~\cite{schmitt2016self,bottcher2019dilute,chomaz2016quantum}. These beyond mean-field phenomena highlight the significant effect of quantum fluctuations in dBECs, often described by the Lee-Huang-Yang (LHY) correction~\cite{lee1957many,lee1957eigenvalues,lima2011quantum,lima2012beyond,boudjemaa2015theory,aybar2019temperature,ozturk2020temperature}. Moreover, beyond the single droplet state observed in free space~\cite{schmitt2016self,Barbut2016PRL,bottcher2019dilute,chomaz2016quantum,baillie2016self,mishra2020self,schmidt2022self,ghosh2022droplet,wachtler2016quantum,bottcher2020new,chomaz2022dipolar}, the interplay between dipole-dipole interaction, quantum fluctuations, and external confinement can give rise to a diverse range of novel states in a single-component dBEC, including droplet arrays~\cite{wachtler2016quantum,bottcher2020new,chomaz2022dipolar,baillie2018droplet,hertkorn2019fate,smith2023supersolidity,chomaz2019long,natale2019excitation,norcia2022can,bottcher2019transient,ghosh2022droplet,poli2023glitches,mukherjee2023supersolid,sohmen2021birth,norcia2021two,bland2022two,halder2022control,sindik2022creation,hertkorn2021supersolidity,tanzi2019observation} and supersolids exhibiting various symmetries~\cite{blakie2020supersolidity,schmidt2022self,hertkorn2021pattern,tanzi2019observation,ripley2023two,wenzel2017striped,gallemi2022superfluid,zhang2023variational,zhang2021phases,schmidt2022self,hertkorn2021pattern,ripley2023two,gallemi2022superfluid,zhang2023variational,zhang2019supersoslidity,zhang2021phases,gallemi2022superfluid,zhang2023variational,zhang2021phases}.

The recent experimental realization of dipolar mixtures~\cite{trautmann2018dipolar} has opened avenues for investigating intriguing beyond mean-field physics in multi-component systems with long-range dipole-dipole interactions. Generally, the presence of multiple species of atoms offers additional interaction channels, potentially giving rise to a variety of unexpected phenomena. Unlike non-dipolar mixtures where collapse can also be suppressed by quantum fluctuations~\cite{petrov2015quantum,petrov2016ultradilute,tengstrand2019rotating,dong2022istable,guo2021lee,cikojeviifmmode2019universality,gautam2019self,hu2020microscopic,filatrella2014domain,hu2020consistent,cabrera2018quantum}, the interplay between inter- and intra-component long-range dipole-dipole interactions, combined with quantum fluctuations, can result in more intricate behaviors, including the tunable miscibility of two-component droplets~\cite{boudjemaa2018fluctuations,bisset2021quantum,smith2021quantum,smith2021approximate,lee2021miscibility,arazo2023self} and the emergence of supersolids with peculiar symmetries~\cite{bland2022alternating,arazo2023self,lee2024excitations,li2022long,halder2023two,halder2023induced,scheiermann2023catalyzation,zhang2024matastable}. Furthermore, the polarization orientation of dipolar mixtures introduces a novel degree of freedom that can be exploited to explore richer phenomena arising from the competition between anisotropic long-range interactions and quantum fluctuations. For instance, in contrast to the droplet state observed in dipolar mixtures with parallel polarization directions~\cite{boudjemaa2018fluctuations,bisset2021quantum,smith2021quantum,smith2021approximate}, theoretical predictions suggest the formation of droplet clusters when the polarization directions are antiparallel in a two-component dBEC~\cite{lee2021miscibility,bland2022alternating,arazo2023self,lee2024excitations}.

While remarkable phenomena such as tunable miscibility and droplet clusters have been predicted in two-component dipolar mixtures with both parallel and antiparallel polarizations, there remains a gap in understanding the related physics when the polarization orientations of the two species are adjusted beyond parallel and antiparallel configurations. Therefore, we aim to delve deeper into the effect of the polarization angle by investigating the corresponding ground-state behaviors of a binary dipolar mixture. To this end, we initially examine the dipolar mixture in free space and observe that both components favour a single droplet state. In comparison with the parallel polarization case, we identify a miscible-immiscible transition between the two species, wherein the critical point depends not only on the inter-component contact interaction but also on the angle between the polarization directions. Moreover, in the immiscible case, the mixture forms two individual droplets composed of each component, which are end-to-end linked with the same angle. Furthermore, upon considering the dipolar mixture confined in an external trap, we uncover that the ground state is altered by multi-droplet states separated by each component alternately in the immiscible regime. Notably, the number of split droplets is primarily controlled by the polarization angle and remains independent of the inter-component contact interaction.

The rest of this paper is organized as follows. In Sec.~\ref{dipolar_mixture_with_arbitrary_angle}, we introduce the binary dipolar mixture system considered here as well as the corresponding theoretical model. In Sec.~\ref{self_bound_ground_state_in_free_space}, we discuss the self-bound droplet states in free space. In Sec.~\ref{phase_diagram}, we present the phase diagram of ground states in the presence of a trapping potential. In Sec.~\ref{quench_dynamics}, we explore the quench dynamics of the confined dipolar mixture. Sec.~\ref{conclusion} provides a conclusion.

\section{DIPOLAR MIXTURE With ARBITRARY Polarization Orientations}\label{dipolar_mixture_with_arbitrary_angle}

\begin{figure}[!b]
	\centering
	\includegraphics[width=0.5\columnwidth]{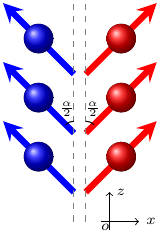}
	\caption{Two-component dipolar mixtures with an arbitrary angle $\alpha$ between their dipole orientations. Red and blue represent component 1 and component 2, respectively. }
	\label{DBEC_alpha}
\end{figure}

We consider a dipolar mixture composed of two species of atoms, and assume the atoms of the two components are polarized in different directions with a relative angle $\alpha$ ($0 \leqslant \alpha \leqslant \pi$) as depicted in Fig.~\ref{DBEC_alpha}. If $\alpha=0$, it reduces to the usual scenario of parallel polarizations, which has been extensively discussed in the typical Dy-Dy as well as Dy-Er mixtures ~\cite{boudjemaa2018fluctuations,bisset2021quantum,smith2021quantum,smith2021approximate,halder2023two,halder2023induced,scheiermann2023catalyzation,li2022long,zhang2024matastable}, while $\alpha=\pi$ corresponds to the case of antiparallel polarizations~\cite{lee2021miscibility,bland2022alternating,arazo2023self,lee2024excitations}.
For the sake of analysis, we assume that the two species are both polarized in the $xz$-plane, and the angle between $z$-axis and the polarization direction of each component is ${\alpha}/{2}$ (see Fig.~\ref{DBEC_alpha}).
Such a setting can be realized by polarizing the two components in their respective directions under strong magnetic fields, and then place them together. Let us take the Dy-Dy mixture as an example, one of the two species can be prepared in the state $\ket{m=8}$, while the other component is initialized in a superposition of the hyperfine spin states~\cite{lee2021miscibility,Ueda2012Spinor}. Despite that the stability of this configuration suffers from dipolar relaxation, which is likely to be suppressed by the development of control techniques~\cite{RelaxationPRA,chalopin2020nphys,barral2023dipolar}, we would like to pursue a deeper understanding of the effect of the relative angle between the polarization orientations in such a scenario.

In contrast to the configuration considered in Ref.~\cite{lee2021miscibility}, where the dipole magnitudes of the two species continuously vary in a wide range and thus corresponds to either parallel or antiparallel cases, the dipole magnitudes of the two components are fixed in our setting, while the dipole orientation is a continuous variable. In such a dipolar condensate mixture, the atoms are subject to long-range dipole-dipole interactions, which can be characterized by~\cite{lahaye2009physics}
\begin{equation}
	V^{\sigma \sigma'}_{\mathrm{dd}}(\mathbf{r})=\frac{\mu_0}{4\pi}\frac{\boldsymbol{\mu}_{\sigma} \cdot\boldsymbol{\mu}_{\sigma'}-3\left(\boldsymbol{\mu}_{\sigma}\cdot\hat{\mathbf{r}}\right)\left(\boldsymbol{\mu}_{\sigma'}\cdot\hat{\mathbf{r}}\right)}{r^3},
\end{equation}
where the two species are indexed by $\sigma,\sigma'=1,2$, $\mu_0$ represents the permeability of vacuum, $\boldsymbol{\mu}_{\sigma}$ denotes the magnetic dipole moment of each component, and $\mathbf{r}$ is the displacement between two separated particles. Hereafter, our discussion will focus on the Dy-Dy mixture, i.e., $\left|\boldsymbol{\mu}_1\right|=\left|\boldsymbol{\mu}_2\right|=\mu=9.93\mu_{\mathrm{B}}$~\cite{bottcher2019dilute,schmitt2016self}, with $\mu_{\mathrm{B}}$ being the Bohr magneton. Subsequently, one can define the dipolar length $a_{\mathrm{dd}}=\frac{\mu_0\mu^2M}{12\pi\hbar^2}\approx 131a_0$~\cite{bottcher2019dilute,schmitt2016self}, where $a_0$ is the Bohr radius and $M$ represents the atomic mass of dysprosium. In the coordinate setting illustrated in Fig.~\ref{DBEC_alpha}, the intra- and inter-component dipole-dipole interactions can be rewritten as
\begin{subequations}\label{ddi}
	\begin{align}
		&\frac{V_{\mathrm{dd}}^{11}(\mathbf{r})}{V_0}=\frac{1}{r^3}-3\frac{x^2 \sin^2\frac{\alpha}{2}+z^2 \cos^2\frac{\alpha}{2}+xz \sin\alpha}{r^5}\\
		&\frac{V_{\mathrm{dd}}^{22}(\mathbf{r})}{V_0}=\frac{1}{r^3}-3\frac{x^2\sin^2\frac{\alpha}{2}+z^2\cos^2\frac{\alpha}{2}-xz\sin\alpha}{r^5}\\
		&\frac{V_{\mathrm{dd}}^{12}(\mathbf{r})}{V_0}=\frac{\cos\alpha}{r^3}-3\frac{z^2 \cos^2\frac{\alpha}{2}-x^2 \sin^2\frac{\alpha}{2} }{r^5}
	\end{align}	
\end{subequations}
with $V_0={3\hbar^2 a_{\mathrm{dd}}}/{M}$.

\begin{figure}[!t]
	\centering
	\includegraphics[width=1\columnwidth]{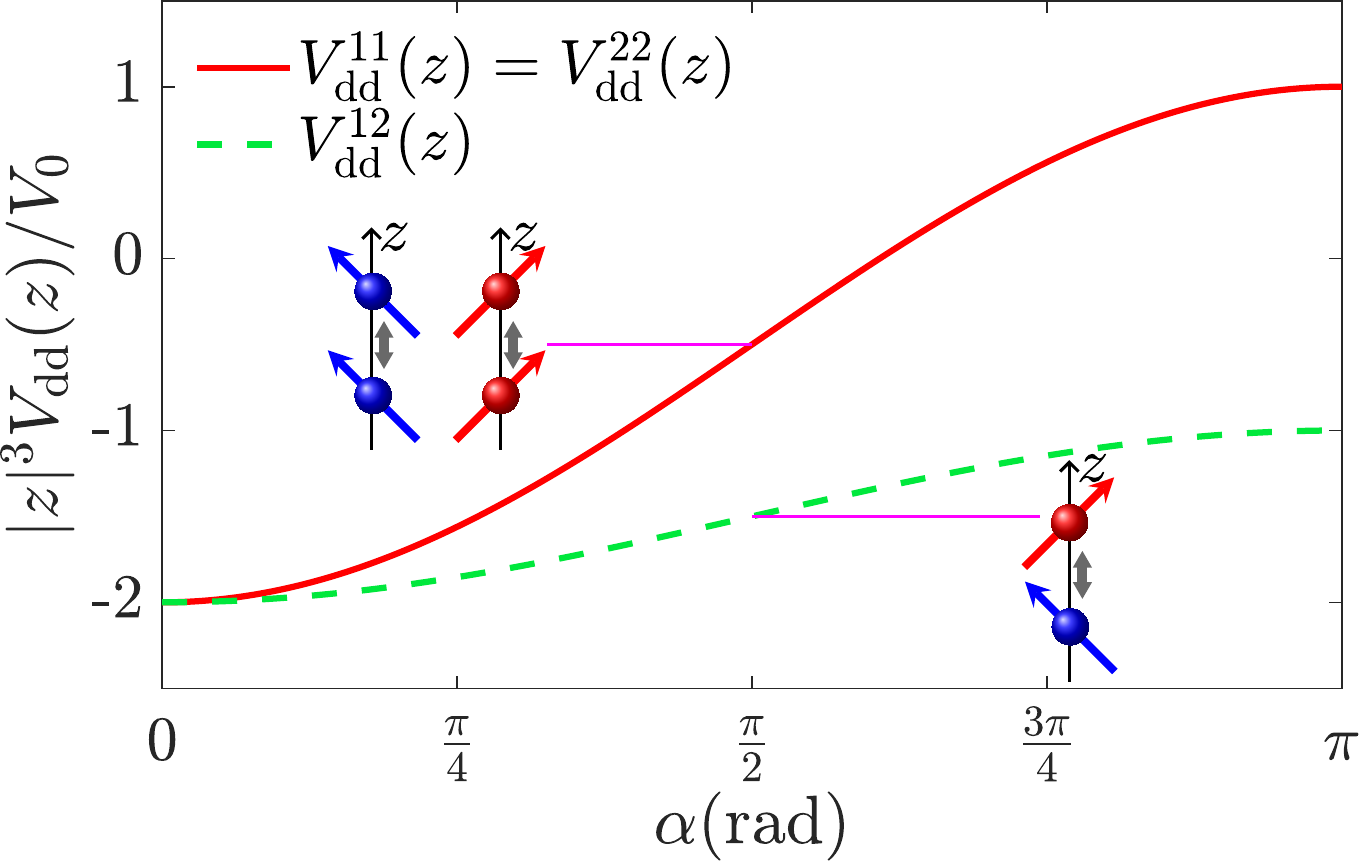}
	\caption{The variation of the intra- (red) and inter-component (green dashed) interaction along $z$-axis $V_{\mathrm{dd}}(z)$ vs $\alpha$. }
	\label{dipole_z}
\end{figure}

For the situation where two particles are aligned along $z$ direction (i.e., $r=|z|$), these dipole-dipole interactions simplify to
\begin{subequations}
	\begin{align}
		V_{\mathrm{dd}}^{11}(z)&=V_{\mathrm{dd}}^{22}(z)=\frac{V_0}{|z|^3}\left(1-3\cos^2\frac{\alpha}{2}\right),\\
		V_{\mathrm{dd}}^{12}(z)&=-\frac{V_0}{|z|^3}\left(1+\cos^2\frac{\alpha}{2}\right).
	\end{align}
\end{subequations}
Fig.~\ref{dipole_z} illustrates the variation of the above intra- as well as inter-component dipole-dipole interactions with respect to the angle $\alpha$. It is noteworthy that the intra-component interaction $V_{\mathrm{dd}}^{\sigma \sigma}(z)$ switches between attraction and repulsion as $\alpha$ increases from 0 to $\pi$, whereas the inter-component interaction $V_{\mathrm{dd}}^{12}(z)$ retains attractive throughout. Due to such attractive long-range interaction along the $z$ direction, the two species are closely bound to each other even in the immiscible scenario, forming either V-type binary droplets in free space or multi-droplet arrays with a zig-zag profile in the presence of a trapping potential, the details of which will be discussed in Sec.~\ref{self_bound_ground_state_in_free_space} and \ref{ground_states_with_a_trap}.

To evaluate the dipolar potential, it is beneficial to express the dipole-dipole interactions in Eq.~(\ref{ddi}) in momentum space via Fourier transformation as below,

	\begin{subequations}
		\begin{align}
			\tilde{V}_{\mathrm{dd}}^{11}(\mathbf{k})&=4\pi V_0  \left(\frac{\left(k_z\cos\frac{\alpha}{2}+k_x\sin\frac{\alpha}{2}\right)^2}{k^2}-\frac{1}{3}\right),\\
			\tilde{V}_{\mathrm{dd}}^{22}(\mathbf{k})&=4\pi V_0 \left(\frac{\left(k_z\cos\frac{\alpha}{2}-k_x\sin\frac{\alpha}{2}\right)^2}{k^2}-\frac{1}{3}\right),\\
			\tilde{V}_{\mathrm{dd}}^{12}(\mathbf{k})&=4\pi V_0 \left(\frac{k_z^2\cos^2\frac{\alpha}{2}-k_x^2\sin^2\frac{\alpha}{2}}{k^2}-\frac{\cos\alpha}{3}\right),
		\end{align}
	\end{subequations}
which is a standard technique to simplify the convolution calculation [see Eq.~(\ref{GPE})].

Based on the above theoretical discussion about the dipole-dipole interactions in such a dipolar mixture with an arbitrary angle between the polarization orientations of the two species, the dynamics of this two-component quantum gases can be described by the following extended Gross-Pitaevskii equation (EGPE)~\cite{bisset2021quantum,smith2021quantum,smith2021approximate}
\begin{equation}
	\begin{split}
		&i\hbar\frac{\partial}{\partial t}\Psi_{\sigma}\left(\mathbf{r},t\right)=\bigg[-\frac{\hbar^2\nabla^2}{2M}+U(\mathbf{r})+
		\sum_{\sigma '}g_{\sigma\sigma '}\left|\Psi_{\sigma '}\left(\mathbf{r},t\right)\right|^2\\
		&+\sum_{\sigma '}\int \mathrm{d}^3 \mathbf{r}' V_{\mathrm{dd}}^{\sigma\sigma '}\left(\mathbf{r}-\mathbf{r}'\right)\left|\Psi_{\sigma '}\left(\mathbf{r}',t\right)\right|^2+\mu_{\mathrm{LHY}}^{(\sigma)}
		\bigg]\Psi_{\sigma}\left(\mathbf{r},t\right)
	\end{split}
	\label{GPE}
\end{equation}
where $\Psi_{\sigma}\left(\mathbf{r},t\right)$ ($\sigma=1,2$) represents the condensed wave function of each component and is normalized to the particle number $\int \left|\Psi_{\sigma}\left(\mathbf{r},t\right)\right|^2 \mathrm{d}^3 \mathbf{r}=N_{\sigma}$, and $g_{\sigma\sigma '}={4\pi\hbar^2 a_{s}^{\sigma\sigma '}}/{M}$ with $a_{s}^{\sigma\sigma '}$ being the $s$-wave scattering length. $U(\mathbf{r})$ denotes the external trapping potential. $\mu_{\mathrm{LHY}}^{(\sigma)}=\frac{\delta E_{\mathrm{LHY}}}{\delta n_\sigma}$ is the LHY correction describing the effect of quantum fluctuations, where $n_\sigma=\left|\Psi_{\sigma}\left(\mathbf{r},t\right)\right|^2$, and
\begin{equation}
	\begin{split}
		E_{\mathrm{LHY}}=&\frac{1}{30\sqrt{2}\pi^3}\left(\frac{M}{\hbar^2}\right)^{\frac{3}{2}}\\
		&\int_{0}^{\pi}\mathrm{d}\theta_k\int_{0}^{2\pi}\mathrm{d}\varphi_k \sin\theta_k \sum_{\lambda=\pm}V_{\lambda}^{\frac{5}{2}}\left(\theta_k,\varphi_k\right)
	\end{split}
	\label{E_lhy}
\end{equation}
with $\theta_k,\varphi_k$ being the spherical coordinates in $\mathbf{k}$-space and
\begin{equation}
	\begin{split}
		V_{\pm}\left(\theta_k,\varphi_k\right)&=\eta_{11}n_1+\eta_{22}n_2\\
		&\pm \sqrt{\left(\eta_{11}n_1-\eta_{22}n_2\right)^2+4\eta_{12}^2 n_1 n_2}.
	\end{split}
\end{equation}
Here, we have defined $\eta_{\sigma\sigma '}=g_{\sigma\sigma '}+\tilde{V}_{\mathrm{dd}}^{\sigma\sigma '}(\mathbf{k})$. It is worth to mention that the above LHY correction contains a small imaginary part, which has been neglected in our simulation, similarly to the usual practice in a single-component dipolar gas~\cite{bisset2016ground,Barbut2016PRL,Saito2016MC} (see more details in the appendix).

\section{SELF-BOUND droplet STATES IN FREE SPACE} \label{self_bound_ground_state_in_free_space}

Let us start with the Dy-Dy mixture without any confinement (i.e., $U=0$) and demonstrate the effect of the polarization orientations by exploring the ground states in the situation of equal particle numbers, i.e., $N_1=N_2$. To obtain the ground states we employ the imaginary time evolution method. That is, propagate the EGPE by replacing $t\rightarrow -i t$ and renormalize the wave function after each propagation step. Eventually, it converges to the least damped state. As this algorithm is sensitive to the initial state, it may converge to either the ground state with the lowest energy or a metastable state with local energy minima. Thus, to accurately determine the ground state, it often requires performing imaginary time evolution with various initial states and subsequently identifying the stationary state with the lowest energy.

\begin{figure*}[!t]
	\centering
	\includegraphics[width=0.9\textwidth]{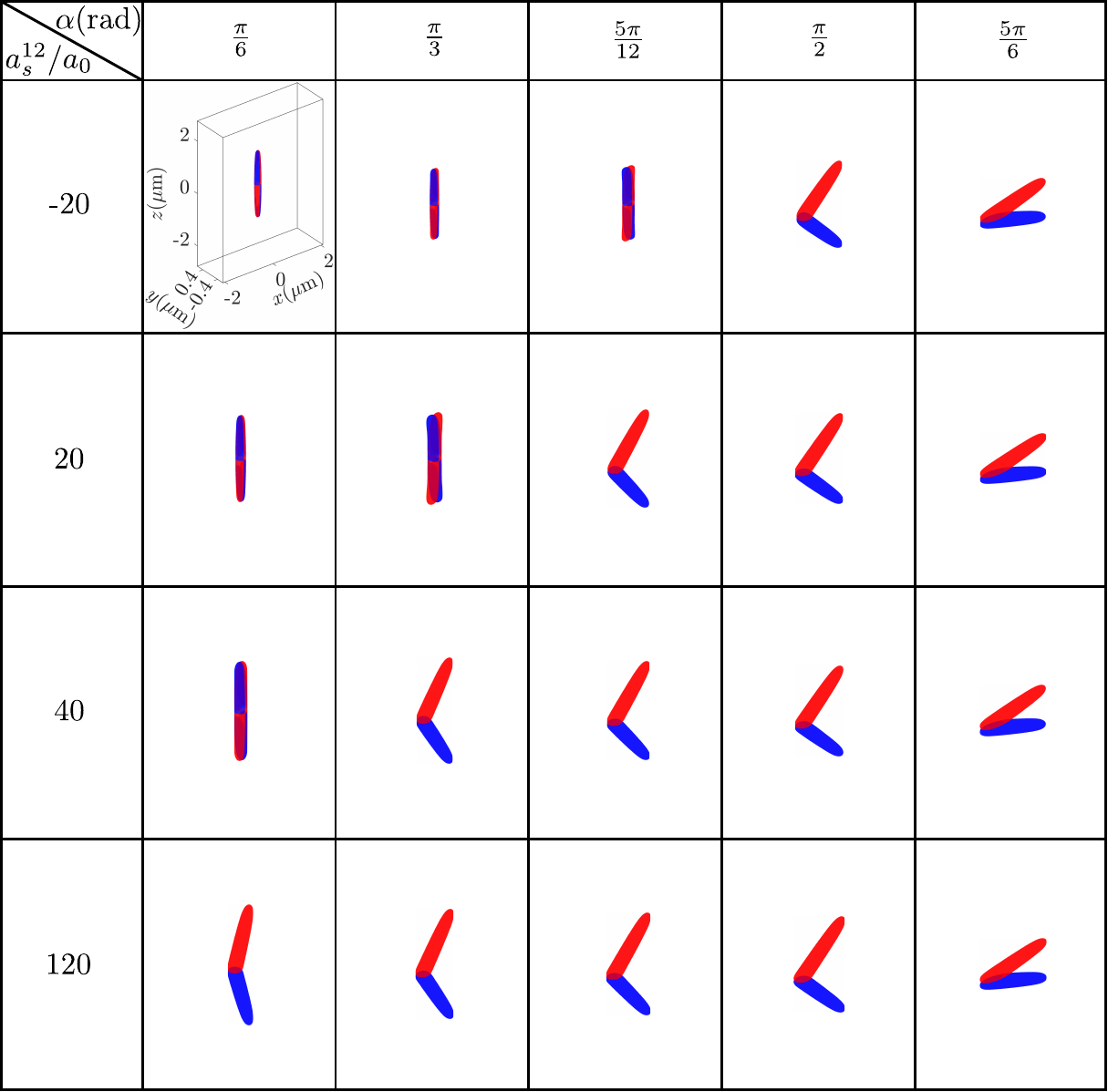}
	\caption{Ground states in free space with fixed atom number $N_1=N_2=1000$. The isosurface corresponds to 5\% of the peak density for each component. All the states are drawn in the same size scale. }
	\label{ground_free}
\end{figure*}

Utilizing this method, we conduct numerical investigations into the ground states of such a balanced dipolar mixture. Fig.~\ref{ground_free} depicts an example illustrating the variation of the ground states concerning the polarization angle $\alpha$ and the inter-component contact interaction $a^{12}_{s}$, while the particle number and the intra-component contact interaction are fixed to $N_1=N_2=1000$ and $a_{s}^{11}=a_{s}^{22}=60a_0$, respectively.

As can be seen from Fig.~\ref{ground_free}, both species exhibit a preference for a single droplet state in free space, accompanied by a miscible-immiscible transition. The miscibility is governed not only by the inter-component $s$-wave scattering length $a^{12}_{s}$ but also by the polarization angle $\alpha$. At small polarization angles and weak inter-component contact interactions, the two components tend to immerse themselves into each other, forming what is known as the miscible droplet state. However, as $a_{s}^{12}$ increases, the dipolar mixture enters the immiscible regime, where the two species spatially separate while remaining connected at the ends of the droplets. This behavior is similar to that observed in the parallel polarization case, where the miscible-immiscible transition is solely governed by $a_{s}^{12}$. Nevertheless, for a fixed $a_{s}^{12}$, the miscible droplet can be transformed into immiscible droplets by adjusting the angle $\alpha$. Moreover, when $\alpha$ is relevantly large (approaching $\pi$), no miscible droplets exist even if the short-range interaction becomes weakly attractive (i.e., $a^{12}_{s}<0$). Additionally, unlike the parallel case, the two droplets are no longer align with each other but instead form a ``V"-like structure. In the immiscible case, the angle of such V-type structure is equal to $\pi-\alpha$; however, it noticeably deviates from $\pi-\alpha$ in the miscible regime due to the strong attraction between the two species.

To quantitatively characterize the angle between the two droplets of each component, we first examine the size of a single droplet $l({\mathbf{n}})$ along an arbitrary direction  ${\mathbf{n}}=\left(\sin\theta\cos\varphi,\sin\theta\sin\varphi,\cos\theta\right)$ as the following
\begin{equation}
	\begin{split}
		l({\mathbf{n}})=&\sqrt{\braket{\left[\left(\mathbf{r}-\braket{\mathbf{r}}\right)\cdot{\mathbf{n}}\right]^2}}\\
		=&\big(\Delta x^2 \sin^2\theta\cos^2\varphi
		+\Delta y^2\sin^2\theta\sin^2\varphi\\
		&+\Delta z^2\cos^2\theta
		+\Delta xy\sin^2\theta\sin 2\varphi\\
		&+\Delta yz\sin 2\theta \sin\varphi
		+\Delta xz\sin 2\theta \cos\varphi \big)^{1/2}
	\end{split}
    \label{droplet_orientation}
\end{equation}
with $\Delta a^2\equiv \braket{a^2}-\braket{a}^2$, $\Delta ab\equiv \braket{ab}-\braket{a}\braket{b}$ ($a,b=x,y,z$), and $\braket{\cdot}$ the average value. And then we define the direction $(\theta_m,\varphi_m)$, along which $l({\mathbf{n}})$ attains its maximum value, as the orientation of the droplet. For the scenario considered here, it is straightforward to notice that the droplets are oriented in the $xz$-plane, and thus  $\varphi_m=0$. In such a case, Eq.~\eqref{droplet_orientation} can be simplified to
\begin{equation}
	l=\sqrt{A\cos\left(2\theta-\phi\right)+B},
	\label{droplet_orientation_xz}
\end{equation}
where $A=\sqrt{\frac{1}{4}(\Delta z^2-\Delta x^2)^2+(\Delta xz)^2}$, $B=\frac{\Delta z^2-\Delta x^2}{2}$, $\phi=\mathrm{arctan2}\left(\Delta xz,\frac{\Delta z^2-\Delta x^2}{2}\right)$.
According to Fig.~\ref{ground_free}, the polar angle orientation of each droplet is in the range of $[0,\pi]$, thus we can identify the following polar angle
\begin{equation}
	\theta_m=\frac{\phi}{2}+{\rm H}\left(-\phi\right)\pi
	\label{eq10}
\end{equation}
with ${\rm H}(x)$ being the Heaviside step function, which allows Eq.~(\ref{droplet_orientation_xz}) to reach its maximum. Subsequently, one can qualitatively define the relative angle between the two droplets formed by each component as $\Delta \theta=|\theta^1_m-\theta^2_m|$.

\begin{figure}[!b]
	\centering
	\includegraphics[width=1\columnwidth]{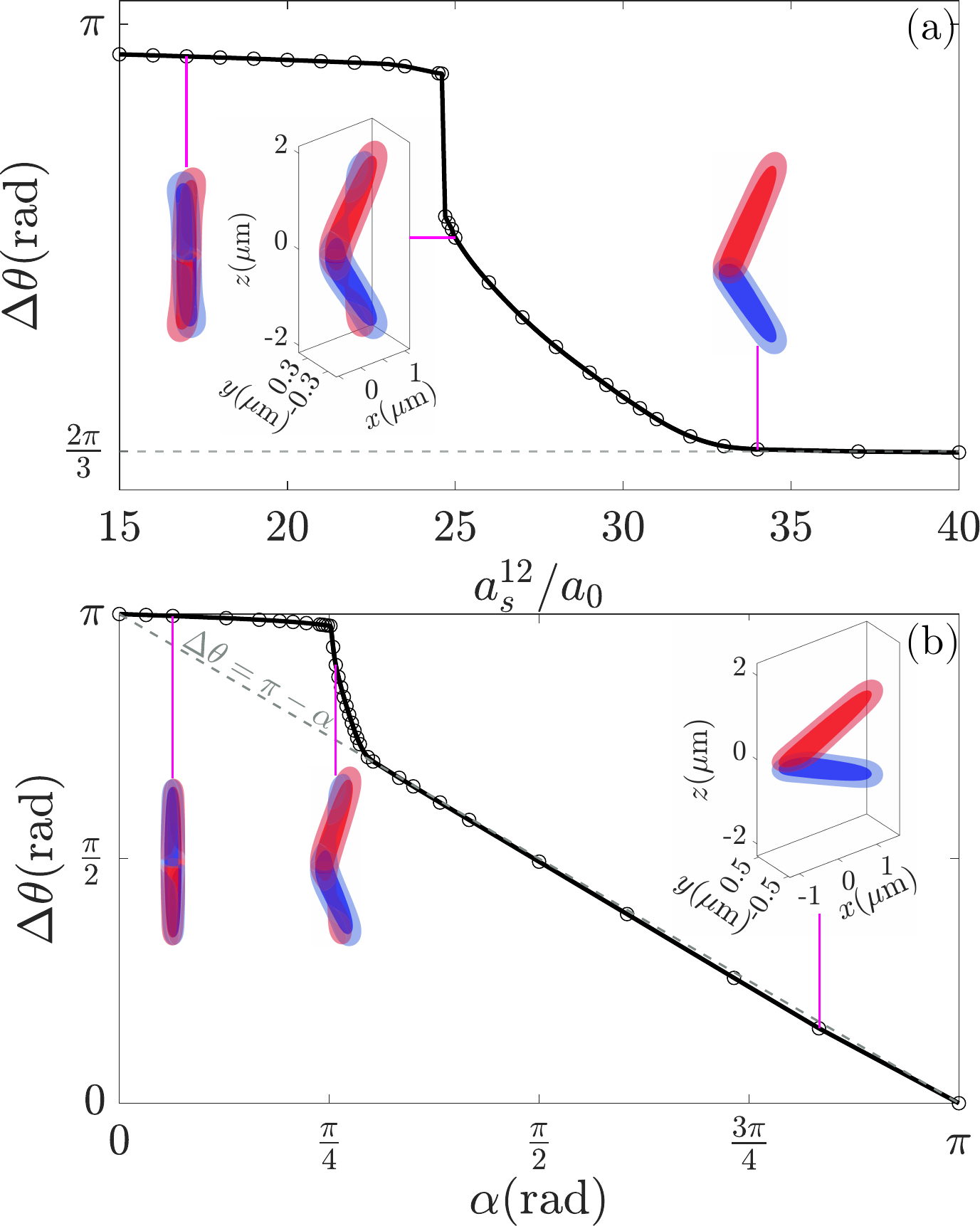}
	\caption{The angle between the two components $\Delta\theta$ vs $a_{s}^{12}$ for $\alpha=\pi/3$ (a) and $\alpha$ for $a^{12}_s=40a_0$ (b). The isosurfaces correspond to 10\% and 0.8\% of the peak density for each component. All the states are drawn in the same size scale.}
	\label{angle}
\end{figure}
Fig.~\ref{angle} shows the variation of the relative angle $\Delta \theta$ between the two droplets across the miscible-immiscible transition point while (a) the polarization angle is fixed at $\alpha=\pi/3$ or (b) the inter-component $s$-wave scattering length is fixed at $a^{12}_s=40a_0$. As can be seen from Fig.~\ref{angle}(a), in the immiscible case at relevantly large $a^{12}_s$, $\Delta \theta$ retains a constant of $2\pi/3$ (i.e., $\pi-\alpha$), which is consistent with the above estimation from Fig.~\ref{ground_free}. As $a_{s}^{12}$ decreases, accompanying with the mixing of the two components, $\Delta \theta$ eventually approaches $\pi$ in the miscible regime, however, the directions of the two components cannot become fully parallel due to the finite polarization angle. The linear relation $\Delta\theta=\pi-\alpha$ for the immiscible droplets is also well justified when tuning the polarization angle $\alpha$ beyond certain threshold for a fixed $a^{12}_s$ as presented in Fig.~\ref{angle}(b). Nevertheless, the droplet angle $\Delta\theta$ converges to $\pi$ at small $\alpha$ where the two species enter miscible regime. From Fig.~\ref{angle}, one can also notice that there is a transition area where the relative angle $\Delta\theta$ takes an intermediate value. This is because that the two components partially bend into each other as displayed by the subplot. The profile of each component in such region is not cigar-shaped, and thus the above Eq.~(\ref{eq10}) is no longer a good estimation for the droplet orientation. Such behavior also manifests that the miscible-immiscible transition is a gradual rather than abrupt variation process. 

\section{GROUND STATES of a confined dipolar mixture}\label{ground_states_with_a_trap}
In this section we turn to the confined Dy-Dy mixture by adding a harmonic trap along $x$-axis, i.e., $U(\mathbf{r})=\frac{1}{2}M\omega^2 x^2$, and explore the ground states in the case of balanced particle numbers, i.e., $N_1=N_2$. In the following, we first present the phase diagram in the subsection~\ref{phase_diagram}, and then discuss the phase transition characteristics in the subsection~\ref{indicators_of _miscible_immiscible_transition}. 

\subsection{Phase Diagram}\label{phase_diagram}
\begin{figure}[!t]
	\centering
	\includegraphics[width=1\columnwidth]{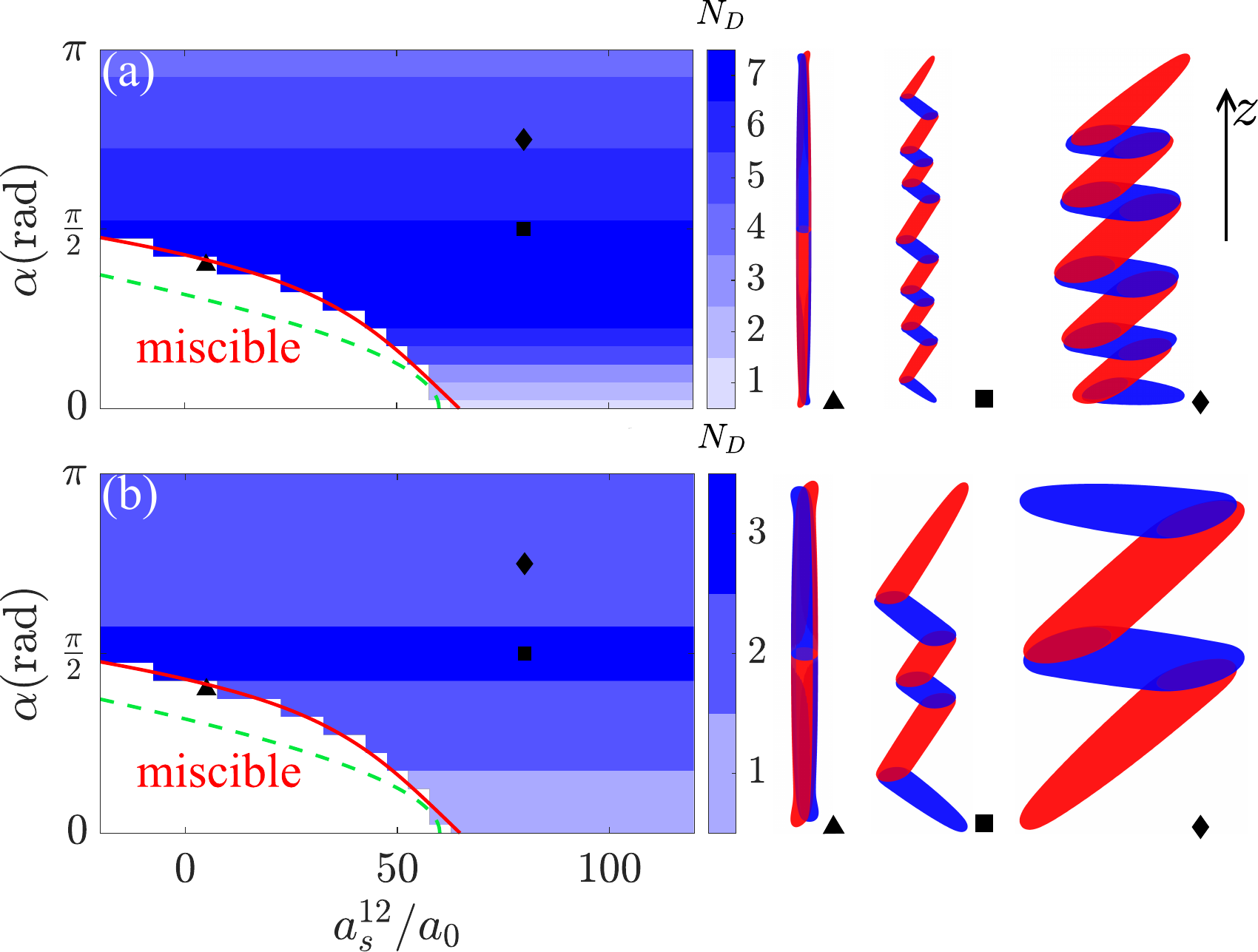}
	\caption{Phase diagram for a dipolar mixture composed of $^{\text{164}}\text{Dy}$ atoms confined in the trap with a frequency $\omega=2\pi\times 400~\mathrm{Hz}$. Here we consider the case of balanced particle numbers, e.g., (a)$N_1=N_2=5\times 10^4$ and (b)$N_1=N_2=10^4$. The red line represents the miscible-immiscible transition boundary fitted from the numerical results, while the green dashed line corresponds to the critical point predicted by Bogoliubov spectrum analysis. $N_D$ refers to the number of separated droplets in one of the two components. The corresponding density profiles of the ground states at the points $\blacktriangle,\blacksquare,\blacklozenge$ are plotted respectively. }
	\label{phase_trapz}
\end{figure}

By fixing the intra-component contact interactions to $a_{s}^{11}=a_{s}^{22}=60a_0$ and tuning the polarization angle $\alpha$ as well as the inter-component $s$-wave scattering length $a^{12}_{s}$, we investigate the ground states of the dipolar mixture confined in the trap with a frequency of $\omega=2\pi\times 400 {\rm Hz}$. Referring to the phase diagram illustrated in Fig.~\ref{phase_trapz}, it is evident that the ground states undergo a miscible-immiscible transition as well. Within the miscible regime, both components form a single droplet bound to each other, exhibiting a small angle similar to that observed in the free space scenario. However, in the immiscible region, the ground state no longer maintains a single droplet for each component. Instead, it adopts a configuration where both species split into multi-droplets, arranged alternately and displaying a typical zig-zag profile reminiscent of similar structures often observed in trapped ions~\cite{Zigzag1,Zigzag2,Zigzag3}.

\begin{figure}[!t]
	\centering
	\includegraphics[width=1\columnwidth]{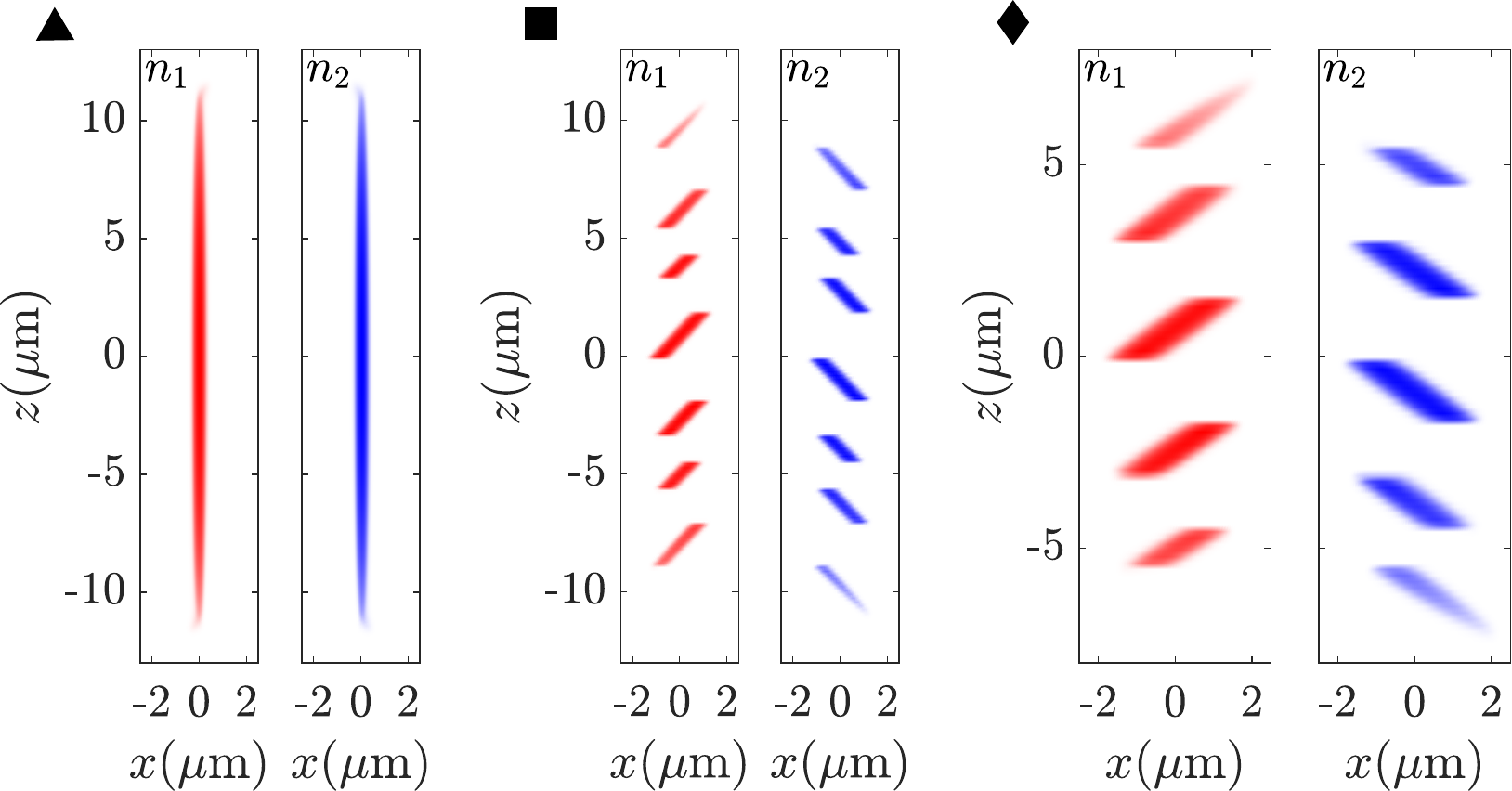}
	\caption{The ground-state density profiles in the $xz$-plane for $N_1=N_2=5\times 10^{4}$ at $a_{s}^{11}=a_{s}^{22}=60a_0$. Here the density has been integrated over $y$-axis, i.e., $n_\sigma (x,z)=\int |\psi_\sigma (\mathbf{r})|^2 {\rm d}y$. The other relevant parameter ($\alpha,\ a^{12}_{s}$) of each state is indicated by the corresponding marks $\blacktriangle,\blacksquare,\blacklozenge$ in Fig.~\ref{phase_trapz}(b).}
	\label{densityxz_xtrap}
\end{figure}

Since the dipolar mixture is confined only in the $x$ direction, the resulting end-to-end linked multi-droplet states are self-bound along the bisector of the polarization angle $\alpha$ (i.e., the longitudinal $z$-direction of the zig-zag droplet chain), owing to the attractive inter-component dipole-dipole interaction (see Fig.~\ref{dipole_z}). If $\alpha < \pi/2$, such immiscible multi-droplet states emerge as the ground state when $a_{s}^{12}$ exceeds a certain threshold (indicated by the red line in Fig.~\ref{phase_trapz}). Conversely, the miscible droplet state is entirely suppressed when $\alpha > \pi/2$ even in the case of weak attractive contact interaction between the two components (i.e., $a^{12}_{s}<0$), which is similar to the behaviour observed in free space. Furthermore, one can also find that the number of the separated droplets of each component remains independent of $a^{12}_{s}$ and varies with the polarization angle $\alpha$. 
To elucidate this behaviour, we analyze the ground-state profiles in the $xz$-plane by integrating the density over the $y$-axis, i.e., $n_\sigma (x,z)=\int |\Psi_\sigma (\mathbf{r})|^2 {\rm d}y$. Fig.~\ref{densityxz_xtrap} exhibits an example for $N_1=N_2=5\times 10^{4}$. In the immiscible regime, the overlap of the two components is minimal, whereas in the miscible case, the two components are nearly superimposed. Additionally, the orientation of the separated droplets of each component is identical to that of the individual droplets in free space (see Sec.~\ref{self_bound_ground_state_in_free_space}), both aligned with their corresponding dipole direction. 
Due to the nearly vanishing density overlap between the two species in the immiscible case, the contribution of the inter-component contact interaction to the total energy is negligible. Consequently, the number of separated droplets keeps constant as $a^{12}_{s}$ varies beyond the threshold. Instead, it is feasible to manipulate the droplet number by tuning the polarization angle. Notably, the droplet number reaches its maximum at $\sim\pi/2$ and decreases as $\alpha$ goes down (up) towards 0 ($\pi$). In addition, by comparing Fig.~\ref{phase_trapz}(a) with (b), it is also worth to notice that the transition point between the miscible and the immiscible regions is not contingent on the total particle number, while the droplet number can be lifted up with increasing particle number.

In Fig.~\ref{phase_trapz}, the miscible-immiscible transition boundary shown by the red line is fitted from the numerical result. Alternatively, such a critical line can also be approximately estimated via the Bogoliubov excitation spectrum~\cite{lee2021miscibility}. Let us assume a homogeneous dipolar mixture in the miscible regime, the Bogoliubov excitation spectrum of which reads

\begin{equation}
	E_{ \pm}^2(\mathbf{k})=\frac{\epsilon_1^2+\epsilon_2^2}{2} \pm \sqrt{\left(\frac{\epsilon_1^2-\epsilon_2^2}{2}\right)^2+n_1 n_2\left(\frac{\eta_{12} \hbar^2 k^2}{M}\right)^2}
	\label{bogoliubov_spectrum}
\end{equation}
where $n_\sigma$ represents the density of each component, and
\begin{equation}
	\epsilon_\sigma^2=\frac{\hbar^2 k^2}{2 M}\left(\frac{\hbar^2 k^2}{2 M}+2 \eta_{\sigma \sigma} n_\sigma\right)
	\label{singledispersion}
\end{equation}
is the single-component dispersion. Here, we have neglected the LHY correction in deriving the above analytical excitation spectrum. While this approach is not entirely rigorous, it serves as a useful tool for qualitatively grasping the underlying physics in a straightforward manner. If the miscible mixture is stable, it is imperative that both the Bogoliubov excitation spectrum Eq.~(\ref{bogoliubov_spectrum}) and the single-component dispersion Eq.~(\ref{singledispersion}) remain purely real as $k\to 0$. To meet this condition, one can readily get the following constraint through some algebra in the case of equal densities $n_1=n_2$:

\begin{equation}
	a_{s}^{12}\leqslant a_{s}^{\sigma \sigma}-\left(1-\cos\alpha\right)a_{\mathrm{dd}}^{\sigma \sigma},
	\label{boundary_condition_alpha}
\end{equation}
beyond which the miscible dipolar mixture is bound to become unstable~\cite{lee2021miscibility}. The miscible-immiscible critical point predicted by Eq.~\eqref{boundary_condition_alpha} is plotted in Fig.~\ref{phase_trapz} as well (see the green dashed line). It presents a good agreement with the numerical result at small $\alpha$, however, a significant deviation between them turns up as the polarization angle increases. Such a discrepancy may, on the one hand, stem from that the dipolar mixture with a finite particle number is not entirely homogeneous, on the other hand, be due to the lack of LHY correction in the Bogoliubov excitation spectrum. Since the LHY correction behaves akin to contact repulsion, which tends to enhance the stability of the miscible mixture, the analytical critical line outlined above would likely be shifted towards the numerical result. Additionally, it is worth to point out that the above miscible-immiscible transition point is independent of the density due to the balanced configuration considered here (i.e., $n_1=n_2$, $a^{11}_s=a^{22}_s$, and $a^{11}_{\rm dd}=a^{22}_{\rm dd}$). This implies that the transition point scarcely varies with the particle number, agreeing with the almost identical boundary shown in Figs.~\ref{phase_trapz}(a) and (b).

\subsection{Miscible-Immiscible Transition}\label{indicators_of _miscible_immiscible_transition}

\begin{figure}[!t]
	\centering
	\includegraphics[width=1\columnwidth]{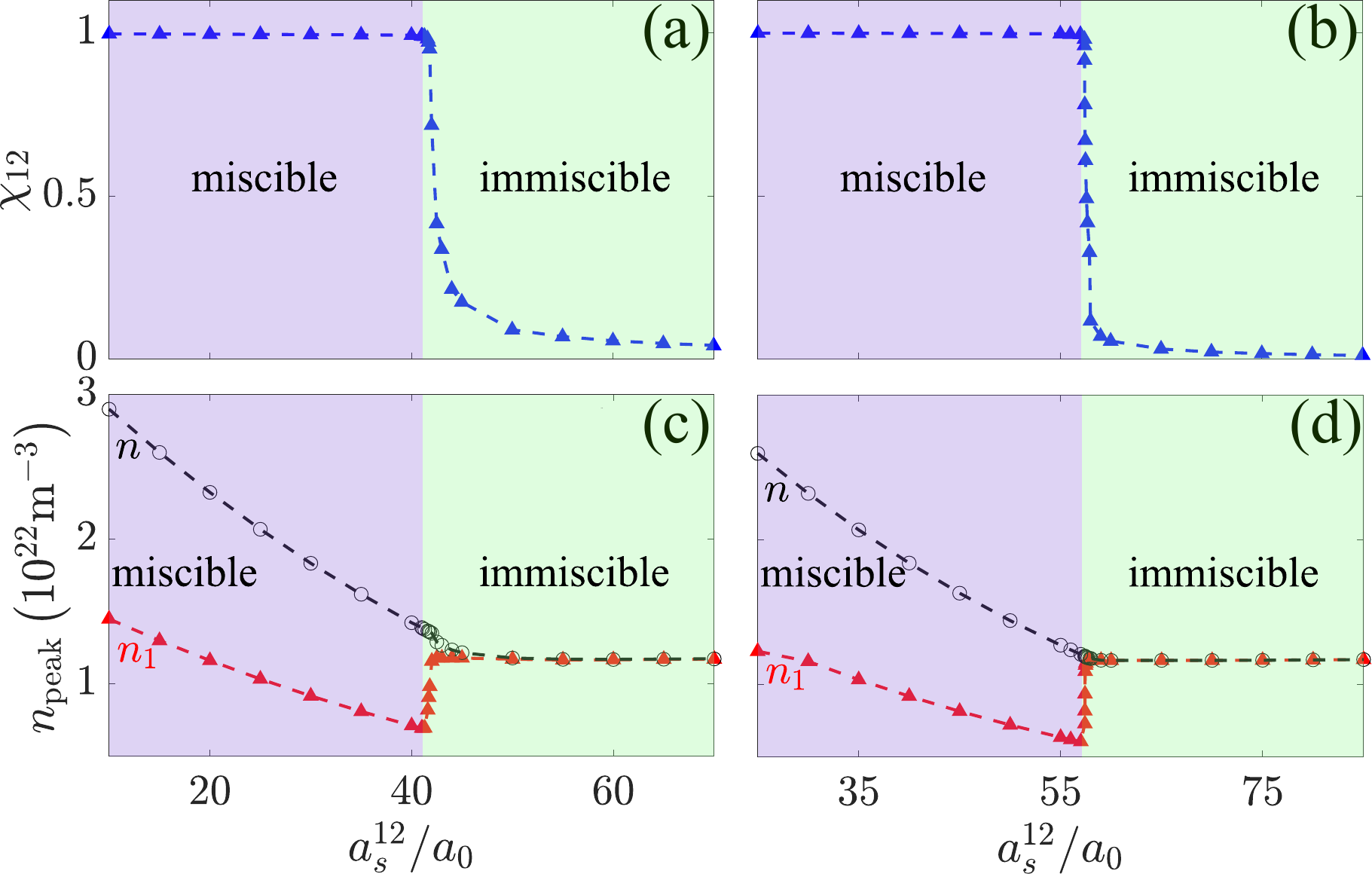}
	\caption{The overlap $\chi_{12}$ (a,b) and the peak densites (c,d) vs $a_{s}^{12}$ for $N_1=N_2=10^{4}$. Here the intra-component $s$-wave scattering length is $a_{s}^{11}=a_{s}^{22}=60a_0$. The polarization angle is fixed at $\alpha={\pi}/{4}$ (a,c) and ${\pi}/{10}$ (b,d), respectively.}
	\label{overlap_npeak}
\end{figure}

As depicted in Fig.~\ref{densityxz_xtrap}, the density profile of the dipolar mixture undergoes a significant change during the miscible-immiscible transition. To quantitatively characterize the overlap between the two species, we employ the following quantity~\cite{smith2021quantum,smith2021approximate}

\begin{equation}
	\chi_{12}=\frac{1}{\sqrt{N_1 N_2}}\int \mathrm{d}^3 \mathbf{r}|\Psi^{\ast}_{1}\Psi_{2}|
\end{equation}
which can serve as an indicator for the miscible-immiscible transition. $\chi_{12}=0$ (1) represents that the two components of the dipolar mixture are completely separated (mixed). As shown in Fig.~\ref{overlap_npeak}(a), for the fixed polarization angle $\alpha=\pi/4$, $\chi_{12}$ retains unity at small $a^{12}_{s}$ in the miscible regime. However, once $a^{12}_{s}$ exceeds the critical point $41a_0$, $\chi_{12}$ gradually decreases and eventually approaches zero as $a^{12}_{s}$ becomes relevantly large, indicating the typical characteristic of the immiscible droplet states. In the case of a smaller polarization angle, for instance, $\alpha=\pi/10$, $\chi_{12}$ experiences a steeper decline around the critical point and rapidly diminishes with increasing $a^{12}_{s}$, which manifests that the transition region between the miscible and fully immiscible regimes contracts as the polarization angle $\alpha$ decreases.

In addition to analyzing the overlap between the two species, we have also investigated the variation of the peak density around the transition point. As displayed in Fig.~\ref{overlap_npeak}(c) and (d), both the total peak density of the mixture, $n_{\rm peak}$, and the peak density of one of the two components (e.g., $n^{\rm peak}_1$) decrease as $a^{12}_{s}$ goes up within the miscible regime. Beyond the critical point, the total peak density smoothly converges to a constant value. In contrast, the peak density of each component initially experiences a rapid rise before saturating to the same constant value as the total peak density. Upon comparison of Fig.~\ref{overlap_npeak}(c,d) with (a,b), one can notice that the miscible-immiscible transition point revealed by the overlap $\chi_{12}$ and by the peak density demonstrates remarkable consistency.

\section{QUENCH DYNAMICS}\label{quench_dynamics}
\begin{figure}[!t]
	\centering
	\includegraphics[width=1\columnwidth]{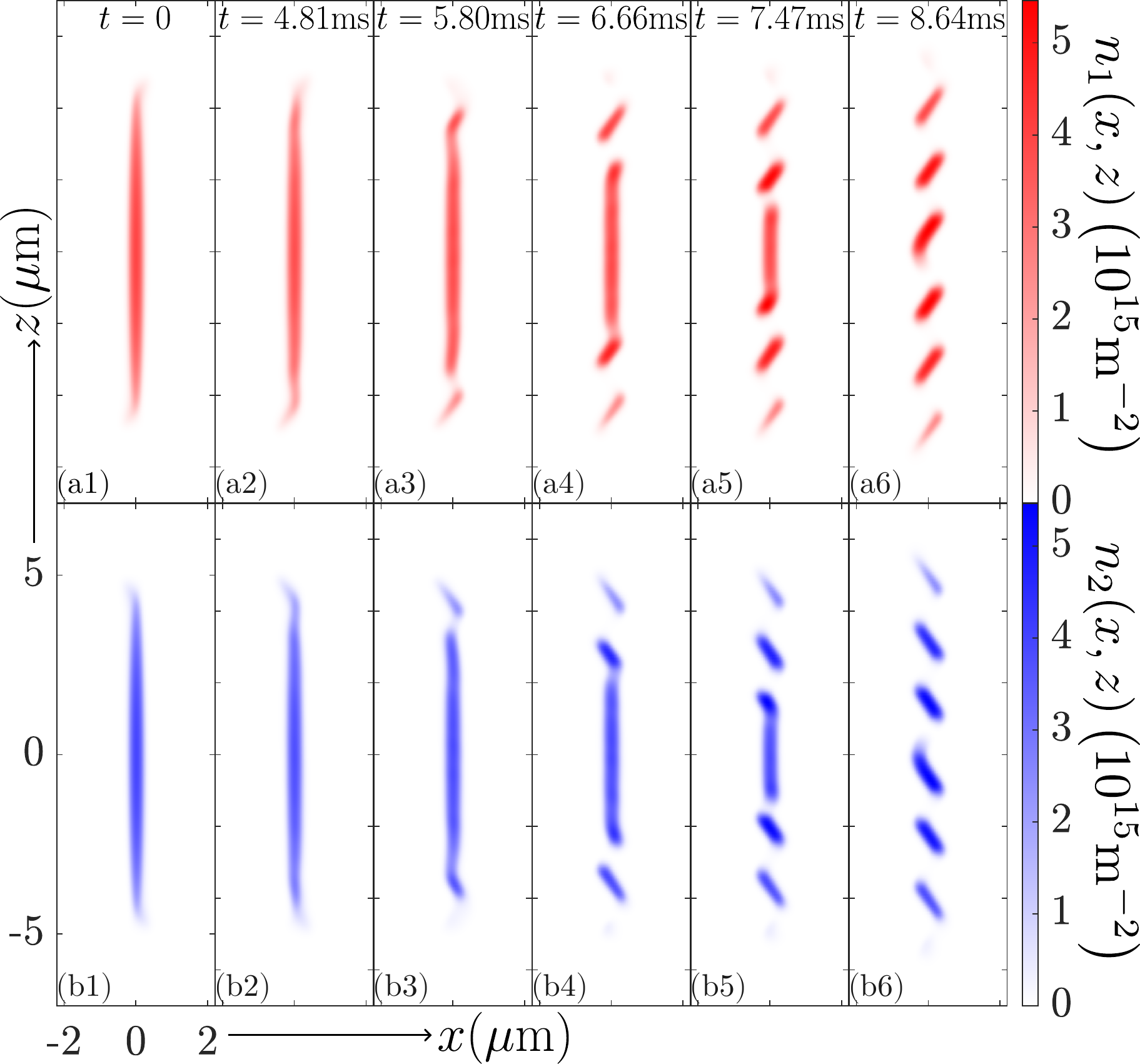}
	\caption{Snapshots of the density distributions $n_\sigma (x,z)$. The upper and the lower panels represent component 1 and component 2 respectively. }
	\label{dynamics}
\end{figure}

To experimentally observe the transition between different states, one standard approach is first preparing the dipolar mixture in one phase and then letting the system evolve after a sudden change of the relevant parameters. Here we consider the quench dynamics of Dy-Dy mixture by tuning the inter-component scattering length $a_s^{12}$. Fig.~\ref{dynamics} presents an example of the dynamics of the dipolar mixture with $N_1=N_2=10000$ particles at $\alpha=\pi/3$, $a_s^{11}=a_s^{22}=60a_0$, and $\omega=2\pi\times 400\mathrm{Hz}$. The dipolar mixture is initialized in the miscible regime at a small $a_s^{12}=5a_0$, i.e., $\blacktriangle$ in Fig.~\ref{phase_trapz}(b), and then $a_s^{12}$ is abruptly lifted up to $10a_0$ in order to migrate the mixture into the immiscible regime. 

As shown in Fig.~\ref{dynamics}, due to the increase of inter-component repulsion, the initial single droplet state is no longer stable and tends to split into multi-droplets. This gradual fission reaches equilibrium after $\sim 8.64\mathrm{ms}$. In comparison with the ground state composed of two separated sub-droplets at the same point $a_s^{12}=10a_0$ (see Fig.~\ref{phase_trapz}), this equilibrium state presents more domains, which is probably caused by the nonadiabatic evolution across the transition boundary~\cite{smith2021quantum}.

\section{CONCLUSION}\label{conclusion}

In this work, we considered a particle-number-balanced Dy-Dy dipolar mixture, where the angle $\alpha$ between the polarization orientations of the two species ranges from 0 to $\pi$, surpassing the parallel ($\alpha=0$) and anti-parallel ($\alpha=\pi$) cases investigated in prior research~\cite{boudjemaa2018fluctuations,bisset2021quantum,smith2021quantum,smith2021approximate,lee2021miscibility,arazo2023self,bland2022alternating,li2022long,halder2023two,halder2023induced,scheiermann2023catalyzation,zhang2024matastable}. This configuration can be achieved by preparing one of the components in spin superposition states. Although the lifetime of such superposition states in dipolar atoms is typically short due to dipolar relaxation~\cite{chomaz2022dipolar}, recent experiments has shown that the dipolar relaxation effect can be significantly mitigated through subtly design of control manners~\cite{chalopin2020nphys,barral2023dipolar}. Such development offers a promising pathway for realizing our proposal in the near future.

By numerically investigating the ground states of such a dipolar mixture, we unveil emergent phenomena by adjusting the polarization angle. Comparing with the scenario of parallelly polarized dipolar mixtures in free space, we show that varying the polarization angle can induce a miscible-immiscible transition for the dipolar mixture. In addition, a finite polarization angle notably shifts the critical point towards smaller $a^{12}_{s}$. Remarkably, when $\alpha > \pi/2$, even with weakly attractive inter-component contact interaction, the miscible region is entirely suppressed. In the immiscible regime, the droplets formed by each component are connected at one end, displaying a V-type profile with an angle equal to $\pi - \alpha$.

Furthermore, when an external trap is present, we demonstrate a modification in the ground state within the immiscible scenario, resulting in multi-droplet states. Specifically, the two species segregate from each other and organize into an array of droplets exhibiting a characteristic zig-zag structure. The number of separated droplets for each component attains its maximum at $\alpha=\pi/2$ and remains unchanged with variations in $a^{12}_{\rm s}$. Moreover, we find that the transition boundary between miscible and immiscible phases is unaffected by the total atom number in the particle-number-balanced configuration considered here. In addition, we also present that the dipolar mixture can evolve into equilibrium multi-droplet states after a quench of the inter-component contact interaction.

These emergent phenomena underscore the profound impact of polarization orientations in dipolar mixtures, introducing a fresh avenue for exploring exotic states of matter in dipolar quantum gases. Investigating novel phenomena associated with polarization orientations in different scenarios, such as Dy-Er mixtures~\cite{trautmann2018dipolar,smith2021quantum} or unbalanced dipolar mixtures, promises to be an intriguing task for future research.

\section*{Acknowledgement}
We are grateful to Abid Ali for valuable comments on the manuscript. This work was supported by the National Key Research and Development Program of China (Grant No.: 2021YFA1401700), National Nature Science Foundation of China (Grant No.: 12104359), Shaanxi Academy of Fundamental Sciences (Mathematics, Physics) (Grant No.: 22JSY036). Y.C.Z. acknowledges the support of Xi'an Jiaotong University through the ``Young Top Talents Support Plan" and Basic Research Funding as well as the High-performance Computing Platform of Xi'an Jiaotong University for the computing facilities.

\appendix
\section{IMAGINARY PART OF LHY ENERGY}
To calculate the Lee-Huang-Yang (LHY) correction, we have utilized the local density approximation, a common technique employed in previous studies~\cite{lima2011quantum,lima2012beyond,boudjemaa2015theory,boudjemaa2018fluctuations,aybar2019temperature,ozturk2020temperature,oliva1989density,chou1996bose,giorgini1997thermodynamics}. This approximation allows for the analytical formulation of the LHY correction, offering a straightforward approach to comprehend the primary physics associated with quantum fluctuations. However, the LHY term obtained using this method is inherently complex, possessing a finite imaginary component. Typically, the imaginary part is disregarded, a reasonable practice in single-component dipolar gases~\cite{bisset2016ground,Barbut2016PRL,Saito2016MC}, where the imaginary component is relatively negligible compared to the real part. Nonetheless, it remains uncertain whether neglecting the imaginary component of the LHY corrections is appropriate for dipolar mixtures with arbitrary polarization angles as considered in this study. Hence, we scrutinize the energy of the LHY correction as follows. According to Eq.~\eqref{E_lhy}, the energy contributed by quantum fluctuations can be rewritten as
\begin{equation}
	\begin{split}
		E_{\mathrm{LHY}}=&\frac{ n^{\frac{5}{2}}}{30\sqrt{2}\pi^3}\left(\frac{M}{\hbar^2}\right)^{\frac{3}{2}}\\
		&\int_{0}^{\pi}\mathrm{d}\theta_k\int_{0}^{2\pi}\mathrm{d}\varphi_k \sin\theta_k \sum_{\lambda=\pm}\mathcal{V}_{\lambda}^{\frac{5}{2}}\left(\theta_k,\varphi_k\right)
	\end{split}
\end{equation}
with
\begin{equation}
	\begin{split}
		\mathcal{V}_{\pm}&\left(\theta_k,\varphi_k\right)=\eta_{11}\gamma+\eta_{22}\left(1-\gamma\right)\\
		&\pm\sqrt{\left[\eta_{11}\gamma-\eta_{22}\left(1-\gamma\right)\right]^2+4\eta_{12}^2 \gamma\left(1-\gamma\right)}.
	\end{split}
\end{equation}
Here, $\gamma={n_1}/{n}$ describes the fraction of Component 1 in the total density of $n=n_1+n_2$. Therefore, the ratio of imaginary part to real part of the LHY energy reads 
\begin{equation}
	\frac{\mathrm{Im}\left(E_{\mathrm{LHY}}\right)}{\mathrm{Re}\left(E_{\mathrm{LHY}}\right)}=
	\frac{{\rm Im}[\int_{0}^{\pi}\mathrm{d}\theta_k\int_{0}^{2\pi}\mathrm{d}\varphi_k \sin\theta_k \sum_{\lambda=\pm}\mathcal{V}_{\lambda}^{\frac{5}{2}}\left(\theta_k,\varphi_k\right)]}{{\rm Re}[\int_{0}^{\pi}\mathrm{d}\theta_k\int_{0}^{2\pi}\mathrm{d}\varphi_k \sin\theta_k \sum_{\lambda=\pm}\mathcal{V}_{\lambda}^{\frac{5}{2}}\left(\theta_k,\varphi_k\right)]}
\end{equation}
It is worth noting that the above ratio is entirely independent of the total density but is primarily determined by the occupancy fraction of the two components.

\begin{figure}[!t]
	\centering
	\includegraphics[width=1\columnwidth]{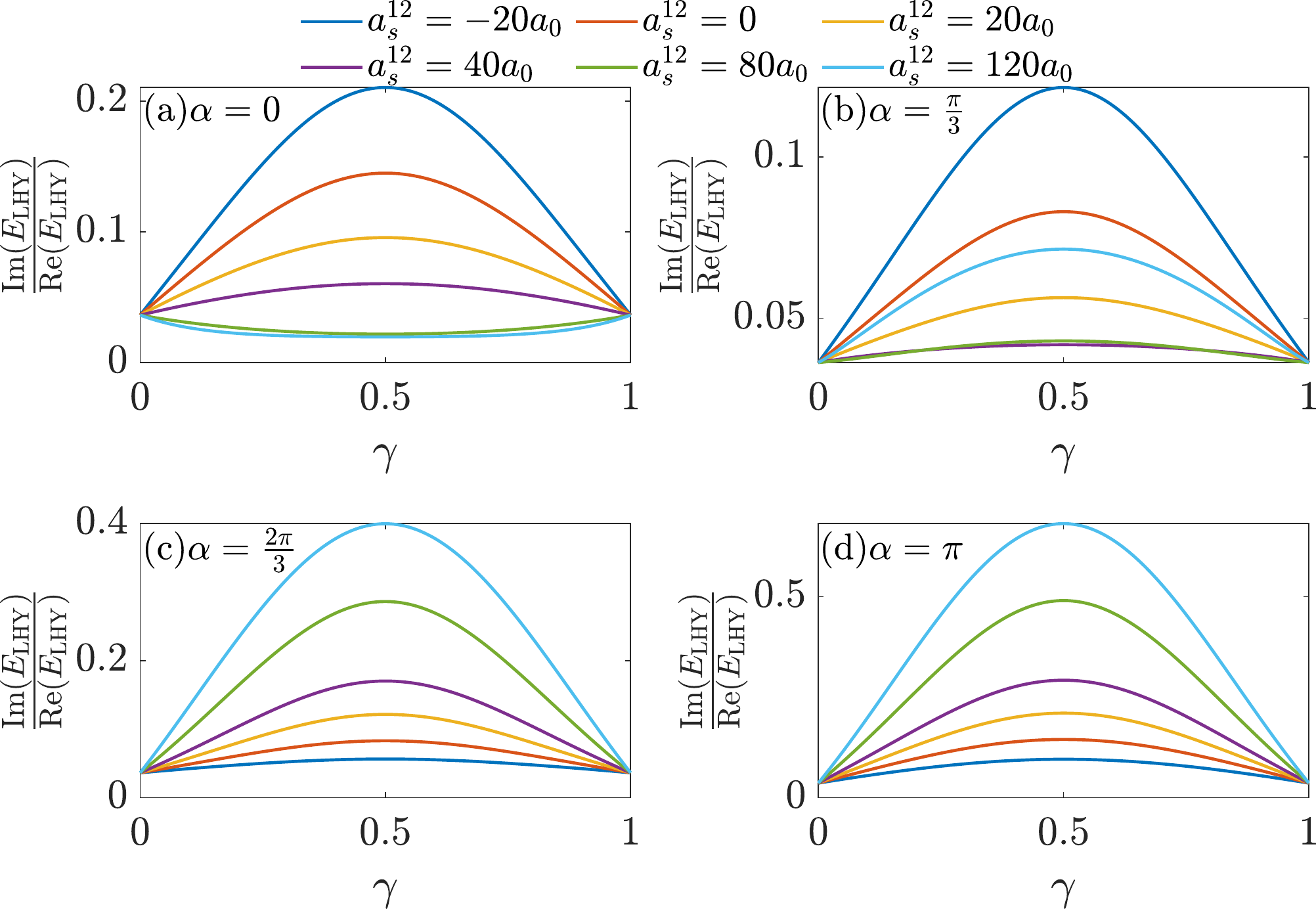}
	\caption{The ratio of imaginary part to real part of the LHY energy for $\alpha=0$ (a), $\pi/3$ (b), $2\pi/3$ (c), and $\pi$ (d). Here, we have fixed the intra-component $s$-wave scattering length at $a_{s}^{11}=a_{s}^{22}=60a_0$. The corresponding value $a^{12}_s$ of each line is indicated by its colour as specified in the top of the figure.}
	\label{LHY_imre}
\end{figure}

In Fig.~\ref{LHY_imre} we plot the ratio of imaginary part to real part of the LHY energy as a function of the occupancy fraction $\gamma$ for different $a^{12}_s$ and $\alpha$. In order to keep consistency with the discussions in Sec.~\ref{self_bound_ground_state_in_free_space} and~\ref{ground_states_with_a_trap}, the intra-component $s$-wave scattering length is fixed to $a_{s}^{11}=a_{s}^{22}=60a_0$. When $\gamma=0$ or $\gamma=1$, it reduces to the situation of a single-component dipolar gas, where the ratio  $\frac{\mathrm{Im}\left(E_{\mathrm{LHY}}\right)}{\mathrm{Re}\left(E_{\mathrm{LHY}}\right)}$ approaches zero. This manifests the reasonability of neglecting the imaginary part in such a case.

However, the amplitude of  $\frac{\mathrm{Im}\left(E_{\mathrm{LHY}}\right)}{\mathrm{Re}\left(E_{\mathrm{LHY}}\right)}$ increases as $\gamma$ approaches 0.5, where the two species of the dipolar mixture have identical density and $|\frac{\mathrm{Im}\left(E_{\mathrm{LHY}}\right)}{\mathrm{Re}\left(E_{\mathrm{LHY}}\right)}|$ attains its maximum. For example, $|\frac{\mathrm{Im}\left(E_{\mathrm{LHY}}\right)}{\mathrm{Re}\left(E_{\mathrm{LHY}}\right)}|_{\rm max}\approx 0.15$ when $\alpha=0$ and $a^{12}_s=0$. 
Clearly, it is not entirely appropriate to ignore the imaginary part in this scenario. Nevertheless, the peak value of $|\frac{\mathrm{Im}\left(E_{\mathrm{LHY}}\right)}{\mathrm{Re}\left(E_{\mathrm{LHY}}\right)}|$ drops with increasing $a^{12}_s$ and becomes less than 0.05 when $a^{12}_s>40a_0$. Thus, it remains reasonable to neglect the imaginary part at a large $a^{12}_s$ for a parallelly polarized dipolar mixture, which has been successfully used to understand the miscible-immiscible transition in binary dipolar mixtures~\cite{smith2021quantum,bisset2021quantum,smith2021approximate}. Similar behaviours can also be observed for a small angle, e.g., $\alpha=\pi/3$ shown in Fig.~\ref{LHY_imre}(b). Moreover, the maximum value of $|\frac{\mathrm{Im}\left(E_{\mathrm{LHY}}\right)}{\mathrm{Re}\left(E_{\mathrm{LHY}}\right)}|$ becomes smaller than that at $\alpha=0$, allowing for the omission of the imaginary part of the LHY correction for both miscible and immiscible binary dipolar mixtures in the case of a small polarization angle.

In contrast, $|\frac{\mathrm{Im}\left(E_{\mathrm{LHY}}\right)}{\mathrm{Re}\left(E_{\mathrm{LHY}}\right)}|$ is significantly lifted up while $\alpha>\pi/2$ as can be seen from Fig.~\ref{LHY_imre}(c,d). In comparison with the case of a small angle discussed above, this ratio becomes increasing with $a^{12}_s$ and can even exceed 0.5 around $\gamma=0.5$ at sufficiently large $a^{12}_s$. It appears that ignoring the imaginary part of the LHY correction at large $\alpha$ beyond $\pi/2$ may no longer be valid. However, under such conditions, the dipolar mixture actually resides in the immiscible regime, where the two components are almost completely separated from each other. That is, in fact, there exist atoms of only one component within each individual sub-droplet (see Fig.~\ref{densityxz_xtrap}), corresponding to either $\gamma=0$ or $\gamma=1$, where $|\frac{\mathrm{Im}\left(E_{\mathrm{LHY}}\right)}{\mathrm{Re}\left(E_{\mathrm{LHY}}\right)}|$ remains negligible. Therefore, we can conclude that neglecting the imaginary part of the LHY correction remains a reasonable approximation in the scenario considered in this work.

\bibliography{mybib}
\end{document}